\documentstyle[12pt]{article}
\begin{document}
\null\vskip1cm
\centerline{\Large\bf A q-Deformed Schr\" odinger Equation}
\vskip1.52cm
\centerline{{\large M. Micu}
\footnote{E-mail address:~micum@theor1.theory.nipne.ro}}
\centerline{Department of Theoretical Physics}
\centerline{ Horia Hulubei Institute of Physics and Nuclear
Engineering}
\centerline{  POB MG-6, Bucharest, 
76900 Romania}
\centerline{and}
\centerline{Joint Institute of Nuclear Research, Dubna, Russia}
\vskip2cm
{\bf Abstract} We found hermitian realizations of the position vector
$\vec{r}$, the angular momentum $\vec{\Lambda}$ and the linear momentum
$\vec{p}$, all  behaving like vectors under the $su_q(2)$
algebra, generated by $L_0$ and $L_\pm$. They are used to introduce 
a $q$-deformed Schr\" odinger equation. Its 
solutions for the particular cases of the Coulomb and the harmonic 
oscillator potentials are given and briefly discussed.

\newpage
{\bf 1. Introduction}
\vskip0.5cm
The general framework of the present study is the theory of
quantum $su_q(2)$ algebra which has been the subject of extensive
developments.
Our purpose is to derive a q-deformed
Schr\" odinger equation invariant under the $su_q(2)$ algebra.
Here we discuss the case of spinless particles.
So far, a general procedure 
(see for example \cite{1}-\cite{3}) was to write down the Hamiltonian
in spherical coordinates and replace the $su(2)$ Casimir operator
$C = \vec{L}^2$ by $C_q$ + $f(q)$ where $q$ is the deformation
parameter, $C_q$ the Casimir
operator of the $su_q(2)$ algebra and $f(q)$ an arbitrary 
function with the property $f(q) \rightarrow 0$ when $q \rightarrow 1$.
Of course this method introduces arbitrariness through the function $f$
and sometimes anomalies as for example a bound spectrum \cite{2}  
for the free Hamiltonian. Here we aim at removing such kind 
of arbitrariness and anomalies.\par  
The novelty of our study is that we search for hermitian 
realizations of the position, momentum and angular momentum
operators behaving as {\it vectors} 
with respect to 
$su(2)_q$ algebra, generated by the operators $L_0$ and $L_{\pm}$.
We shall show that the angular momentum entering
the expression of the Hamiltonian has components
$\Lambda_0$ and $\Lambda_{\pm}$, different from $L_0$ and $L_{\pm}$.
This leads to a proper behaviour of the free Hamiltonian. Here we consider
two cases of central potentials: the harmonic oscillator and
the Coulomb potential. Once the Hamiltonian is constructed
we are able to derive both the spectrum and the eigenfunctions 
in a consistent way for each case.
Our arguments are as follows.\par
The usual quantum mechanics of a point-like particle is constructed from 
two vectors: the position vector $\vec{r}$ and the linear momentum
$\vec{p}=-i \hbar \vec{\nabla}$.
These two vectors are 
used to build all the other quantities, as e.g. the angular momentum, 
the interaction
potentials, etc., according to the classical rules. In general, these 
operators do not commute, 
their commutation relations following from
the commutation relations of $\vec{r}$ and $\vec{p}$. 

In a $q$-deformed quantum mechanics the commutation relations between 
the generators of the $su_q(2)$
algebra, $L_i$, and the position vector $\vec{r}$ are well defined 
inasmuch as $\vec{r}$ is considered a q-tensor of rank one (see next section).
Therefore it is natural to take $r_i$  as the basic quantities from which 
all the others should be built. Then in deriving
a $q$-deformed Schr\" odinger Hamiltonian, invariant
under the $su_q(2)$ algebra, we searched
for a realization of the linear momentum $\vec{p}$ entering the kinetic
energy term. First it was necessary to find a realization
for $\vec{r}$
and for $L_i$ as {\it self adjoint} operators obeying 
commutation relations characteristic to the deformed algebra.
Then we looked for a realization of 
$\vec{p}$ in terms of $\vec{r}$ and of $L_i$. We found that 
$\vec{p}$ can be written as a sum of two terms which are parallel 
and perpendicular to $\vec{r}$ respectively. As discussed 
below, the parallel component of $\vec{p}$ is assumed to
have the simplest possible form and is written as
$-i{\vec{r}\over r^2}~(r{\partial\over\partial r}+1)$
while the perpendicular one must be expressed as a vector product of $\vec{r}$
and of $\vec{\Lambda}$.  

The paper is organized as follows.
Section 2 contains the general commutation relations involving 
the $q$-angular momentum.  We introduce
some quantities having definite
transformation properties with respect to the $su_q(2)$ algebra,
namely the invariants C, ${\rm C}'$ and c and the vector $\vec{\Lambda}$ 
related to $\vec{L}$. 

In the third section we propose a realization of the position vector
$\vec{r}$ and consistently
of the $q$-angular momentum $\vec{L}$, in terms of
spherical coordinates $r,~x_0={\rm cos}\theta$ and $~\varphi$, as for example  
in Refs. \cite{4},\cite{5}.

The realization of the linear momentum $\vec{p}$ is considered in 
the fourth section. We first build the part of $\vec{p}$ perpendicular to
$\vec{r}$, denoted by $\vec{\partial}$. This is achieved by using the cross
product $\vec{r} \times\vec{\Lambda}$. We find that the components of 
$\vec{\partial}$
satisfy the same type of 
commutation relations as the components of $\vec{r}$.

Section 5 introduces the eigenfunctions of the $q$-deformed angular
momentum written as power series of 
$x_0={\rm cos}\theta$. We show that the result is a
generalization of 
the hypergeometric functions $_2{\rm F}_1(a,b,c;{1\over2};x_0^2)$
and  $_2{\rm F}_1(a,b,c;{3\over2};x_0^2)$ which can be related to
the $q$-deformed spherical functions $Y_{lm}(q,x_0,\varphi)$.
Some useful properties and relations satisfied by the eigenfunctions
are proved. 
In the last section two  particular cases of $q$-deformed
Schr\" odinger equation containing a scalar potential are presented: 
the Coulomb and the three dimensional oscillator.
Their eigensolutions are given and 
the removal of the accidental degeneracy is discussed.
 
\vskip1cm
{\bf 2. The {\large$q$}-angular momentum}
\vskip0.5cm
The $su_q(2)$ algebra is generated by
three operators $L_+,~L_0$ and $L_-$, also named the $q$-angular
momentum components. They have the following commutation relations:
$$\left[~L_0~,~L_\pm~\right]~=~\pm~L_\pm,\eqno(1)$$
$$\left[~L_+~,~L_-~\right]~=~\left[2~L_0\right],\eqno(2)$$
where the quantity in square brackets is defined as
$$\left[n\right]~=~{q^n-q^{-n}\over q-q^{-1}}.\eqno(3)$$
In the following we shall introduce quantities
having definite 
transformation properties with respect to
the $su_q(2)$ algebra. They will further be used to build
$q$-scalars and also $q$-vectors,
as for instance, the $q$-linear momentum entering the expression
of the Hamiltonian operator.

First of all we recall that $su_q(2)$ algebra has an invariant
C, called the Casimir operator 
$${\rm C}~=~L_-~L_+~+~\left[L_0\right]~\left[ L_0~+1\right
].\eqno(4)$$
Its eigenvalue associated to a $(2l+1)$-dimensional irreducible
representation is:
$${\rm C}_l=[l]~[l+1].\eqno(5)$$
By definition 
a $q$-vector in this algebra is given by a set of three quantities
$v_k,~k= 0, \pm1$ satisfying the following relations:
$$\left[~L_0~,~v_k~\right]~=~k~v_k,\eqno(6)$$
$$\left(~L_\pm~v_k~-~q^k~v_k~L_\pm~\right)~q^{L_0}~=~\sqrt{[2]}~
v_{k\pm1},\eqno(7)$$
where $v_{\pm2}$ must be set equal to zero in the right-hand side
of equation (7) 
when $k=\pm1$. This definition is a particular case of an irreducible
tensor of rank one (for the general case see e.g. Ref. \cite{6}).   

By comparing the relations (1), (2) with (6), (7) we observe that 
the operators $L_k$ do not represent the components of a $q$-vector.
Such an observation is also pointed out in Ref. \cite{7} in the
context of q-tensor operators for quantum groups.  
The situation is entirely different from the $su(2)$ algebra
where $L_k$ form a vector in the usual sense.
However, one can use the components $L_\pm$ and $L_0$ to
define a new vector  $\vec{\Lambda}$  
in the following manner:
$$\Lambda_{\pm1}~=~\mp 1~\sqrt{\frac{1}{[2]}}~q^{-L_0}~L_\pm,\eqno(8)$$

$$\Lambda_0~=~{1\over[2]}~\left(q~L_+~L_-~-~q^{-1}~L_-~L_+\right).
\eqno(9)$$
It is an easy matter to show that the operators $\Lambda_k$  satisfy the
relations (6) and (7). The vector $\vec{\Lambda}$ will be used in 
Section 4 to construct the transverse part of the linear momentum
$\vec{p}$.

Two $q$-vectors $\vec{u}$ and $\vec{v}$ satisfying equations (6) and (7) 
can be used to build a 
scalar $S$, according to the following definition:
$$S~=~\vec{u}~\vec{v}~=~-{1\over q}~u_1~v_{-1}~+~u_0~v_0~-~q~
u_{-1}~v_{1}.\eqno(10)$$
where the coefficients appearing in the sum  
are the $q$-Clebsch-Gordan coefficients 
${\langle 1 1 0 | m -m 0\rangle}_q$. In this way the scalar product (10)
becomes the ordinary scalar product of $R(3)$ when $q = 1$.  
By introducing a generalization of the cross product, two $q$-vectors
can also be used to build another $q$-vector required by our approach,  
as it will be shown in Section 4. 

In the case $\vec{u}=\vec{v}=\vec{\Lambda}$, the scalar product
$\vec{\Lambda}^2$ defines a second invariant \cite{8}, ${\rm C}'$, which
is not independent of C. The eigenvalue of ${\rm C}'$ is:
$${\rm C}'_l~=~{[2l]\over[2]}~{[2l+2]\over[2]}.\eqno(11)$$
One can also easily prove that there exists 
a third invariant c, defined as
$${\rm c}~=~q^{-2L_0}~+~\lambda~\Lambda_0,\eqno(12)$$
with
$$\lambda~=~q-{1\over q}.\eqno(13)$$
This will be frequently used in order to write the subsequent
formulae in a more compact form.
Its eigenvalue is:
$${\rm c}_l~=~{q^{2l+1}~+~q^{-2l-1}\over[2]}.\eqno(14)$$
It is worth noting that in the limit $q \rightarrow$ 1 both the  
${\rm C}$ and ${\rm C}'$ turn into the Casimir
invariant C=$\vec{L}^2$ of $su(2)$ with the eigenvalue $l(l+1)$,
while c becomes equal to unity. The eigenvalues (11) and (14)
will be used in section 4 to define the action of $\vec{p}^2$
on deformed spherical harmonics.
The results listed in this section are valid for any realization
of the $su_q(2)$ algebra.
\vskip1cm
{\bf 3. The position vector {\large$\vec{r}$} and a realization 
of {\large$L_0$}, 
{\large $L_\pm$}} 
\vskip0.5cm
In the $R_q(3)$ space we define 
the position vector $\vec{r}$ as having three
noncommutative components $r_1,~r_0$ and $r_{-1}$, satisfying the
following relations
$$r_0~r_{\pm1}~=~q^{\mp2}~r_{\pm1}~r_0,\eqno(15)$$
$$r_1~r_{-1}~=~r_{-1}~r_1~+~\lambda~r_0^2.\eqno(16)$$
These equations are similar to eqs. (3.11) of Ref. \cite{1}
They are typical for a noncommutative algebra. 
The scalar quantity $r^2$ defined according to equation (10)
$$r^2~=~\vec{r}~^2~=~-{1\over q}~r_1~r_{-1}~+~r_0^2~-~q~
r_{-1}~r_1\eqno(17)$$
commutes with all $r_i$ and  all $L_i$ of equations (1) and (2), provided 
$r_i$ $( i = 0, \pm1 )$ satisfy the conditions (6) and (7) to be a vector,
which is here the case. For
$q=1$ the scalar $r$ is nothing else but the length of the
position vector $\vec{r}$. We shall keep this meaning for
$q\neq1$ too.

Searching for concrete realizations of $r_i$, $L_0$ and
$L_\pm$, we begin by expressing $L_0$ in spherical coordinates as in 
the $R(3)$ case: 
$$L_0~=~-i~{\partial\over\partial\varphi}.\eqno(18)$$
The next step is to write $\vec{r}$ as a product of $r$ and of a
unit vector $\vec{x}$, depending on angles 
so we have:
$$r_{\pm1}~=~r~x_{\pm1},\eqno(19)$$
$$r_0~=~r~x_0.\eqno(20)$$
It remains now to find a realization of $x_{\pm1}$ in terms of the
azimuthal angle $\varphi$ and of $x_0$, which is in fact equal to
$\cos\theta$, just as in the classical $R(3)$ case. 
>From the relations (16), (17), (19) and (20) one can find
$$x_{1}~x_{-1}~=~-~\frac{1}{[2]}~(1-q^2~x_0^2)~,$$
$$x_{-1}~x_{1}~=~-~\frac{1}{[2]}~(1-q^{-2}~x_0^2).$$
This suggests that the equations (15) and (16) can be satisfied
by simple forms of $x_{1}$ and $x_{-1}$ provided a dilatation
operator $N_0$ is introduced through the commutation relations
$$[~N_0~,~x_0^n~]~=~n~x_0^n,\eqno(21)$$
and having the hermiticity property
$$N_0^+~=~-N_0~-~1.\eqno(22)$$
Then the realization of $x_1$ and $x_{-1}$ satisfying (15) and (16)
turns out to be 
$$x_1~=~-e^{i\varphi}~\sqrt{q\over[2]}~\sqrt{1-q^2~x_0^2}~
q^{2N_0},\eqno(23)$$ 
$$x_{-1}~=~e^{-i\varphi}~\sqrt{1\over[2]q}~\sqrt{1-q^{-2}~x_0^2}~
q^{-2N_0}.\eqno(24)$$ 

Taking now into account the relations (21) and (22) and assuming 
$$x_0^+~=~x_0.\eqno(25)$$ 
we get the expected hermiticity properties for $x_\pm$ as:
$$x_1^+~=~{-1\over q}~x_{-1},\eqno(26)$$
$$x_{-1}^+~=~-q~x_1.\eqno(27)$$
All these arguments allow us to conclude that eqs.(19-25) define
the realization of 
the position vector $\vec{r}$ in the $R_q(3)$ space.

The following step is to search for a realization of the $su_q(2)$
generators. The expressions we propose for $L_+$ and $L_-$ are:
$$L_+~=~\sqrt{[2]}~e^{i\varphi}~{\tilde x}_1^{L_0+1}~{1\over
x_0}~{1-q^{-2N_0}\over 1-q^{-2}}~{\tilde x}_1^{-L_0}~q^{L_0},\eqno(28)$$
$$L_-~=~\sqrt{[2]}~e^{-i\varphi}~{\tilde x}_{-1}^{-L_0+1}~{1\over
x_0}~{1-q^{2N_0}\over 1-q^2}~{\tilde x}_{-1}^{L_0}~q^{L_0},\eqno(29)$$
where ${\tilde x}_{\pm1}={\rm e}^{\mp i\varphi}~x_{\pm1}$ depend on
$x_0$ only. The reason why instead of $x_{\pm1}$ 
we use here ${\tilde x}_{\pm1}$,
where the phase factor has been removed,
is that expressions like 
$x_{\pm1}^{L_0}$ have no meaning,
while ${\tilde x}_{\pm1}^{L_0}$ are well defined
as discussed below equation (30).  
>From equations (18), (28) and (29) we can now construct the 
Casimir operator ${\rm C}$ of equation (4).
Its eigenfunctions are expected to be $q$-spherical functions
as in Ref. \cite{4}. For $q$ = 1 they 
become ordinary spherical harmonics. Therefore they can take the form:
$${\tilde Y}_{lm}(q,x_0,\phi)~=~{\rm e}^{im\varphi}~{\tilde x}_1^m
~\Theta_{lm}(x_0),\eqno(30)$$
where $~\Theta_{lm}(x_0)$ are the q-analogue of the associated Legendre functions.
The functions (30) will be derived and normalized in Section 5. 
 
Concerning the action of $L_+$
we note that ${\tilde x}_1^{-L_0}$ in (28) removes the factor
${\tilde x}_1^m$
in ${\tilde Y}_{lm}(q,x_0,\phi)$. In this way one prevents $q^{-2N_0}$
appearing in equations (28) and (29) from acting on
$\tilde{x}_1$ and producing a troublesome result. The the operator
$q^{-2N_0}$ in $L_+$ acts on $\Theta_{lm}(x_0)$ only. Finally 
e$^{i\varphi}~{\tilde x}_1^{L_0+1}$ recreates the factor $x_1^{m+1}$
required  after it has been eliminated because of
${\tilde x}_1^{-L_0}$.

In the well known $R(3)$ theory of angular momentum a different
mechanism prevents ${\partial\over\partial\theta}$ in $L_+$ from acting on
$x_1^m$: the term given by ${\partial\over\partial\theta}~x_1^m$ is exactly
cancelled out by $i{\rm ctg}\theta{\partial\over\partial\varphi}~x_1^m$, 
so that only the derivative ${\partial\over
\partial\theta}\Theta_{lm}(x_0)$ remains.

It can be verified that the expressions (18), (28) and (29)
satisfy the commutation relations (1) and (2) and hence one can
conclude that they are the realization of the $su_q(2)$
generators in the $R_q(3)$ space. It can also be checked that the
position vector $\vec{r}$, defined by (19-22), behaves indeed as  
a vector in this $su_q(2)$ algebra, since it satisfies the
relations (6) and (7) with $L_\pm$ given by (28) and (29).

\vskip1cm
{\bf 4. The {\large$q$}-linear momentum {\large$\vec{p}$}}
\vskip1cm
In order to write down an expression for the linear momentum
$\vec{p}$, we separate it into a part perpendicular and another one 
parallel to $\vec{x}$. The first one is defined with the
aid of the cross product $\vec{x}\times\vec{\Lambda}$
and the second
one is assumed to have the form $\vec{x}~{1\over
r}~f\left(r{\partial\over\partial r}+1\right)$, 
where $f$ is a function which will be defined in the following.
Then the  components of the transverse part, denoted by $\partial_k$,  
read:
$$\partial_1~=~ q^{-1}~x_1~\Lambda_0~-~q~x_0~\Lambda_1~+~
x_1~c,\eqno(31)$$
$$\partial_0~=~x_1~\Lambda_{-1}~-~\lambda~ x_0~\Lambda_0~-~x_{-1}
~\Lambda_1~+~x_0~c,\eqno(32)$$
$$\partial_{-1}~=~- q~x_{-1}~\Lambda_0~+~q^{-1}~x_0~\Lambda_{-1}~+~
x_{-1}~c,\eqno(33)$$
where $c$ is the invariant defined in equation (12) and
the terms $x_k~c$ have been added to the cross product $\vec{x}\times
\vec{\Lambda}$ in order to ensure the well defined
character with respect to the hermitian conjugation operation
$$\partial^+_k~=~-\left(-{1\over q}\right)^k~\partial_{-k}. \eqno(34)$$
It can be checked that the quantities $\partial_k$ 
form a vector as defined by equations (6) and (7). Moreover they
satisfy the following relations: 
$$\partial_0\partial_{1}~=~q^{-2}~\partial_{1}\partial_0,\eqno(35)$$
$$\partial_0\partial_{-1}~=~q^{2}~\partial_{-1}\partial_0,\eqno(36)$$
$$\partial_1\partial_{-1}~=~\partial_{-1}\partial_1~+~\lambda~
\partial_0^2.\eqno(37)$$
These equations are similar to (15) and (16) satisfied by the 
position vector.
Equation (35) has been directly obtained by commuting $\partial_0$ with
$\partial_1$. Equation (36) is the hermitian conjugate of the
above one. Equation (37) can be obtained either from (35) or
(36) by using the relation (7).

Also, by multiplying equations (31-33) with the corresponding
$x_k$ and taking into account the commutation relations (6) and (7)
one gets:
$$\vec{x}~\vec{\partial}~=~-\vec{\partial}~\vec{x}=c~.\eqno(38)$$
By commuting the invariant $c$ with $\vec{x}$ one finds:
$$\vec{\partial}~=~\lambda^{-2}~\left[c,\vec{x}\right].\eqno(39)$$
Taking now the matrix elements of the last relation one obtains:
$$\left\langle~
l+1~m'~\vert~\vec{\partial}~\vert~l~m\right\rangle~=~{[2l+2]\over[2]}
~\left\langle~l+1~m'~\vert~\vec{x}~\vert~l~m\right\rangle,\eqno(40)$$
$$\left\langle~
l-1~m'~\vert~\vec{\partial}~\vert~l~m\right\rangle~=~-{[2l]\over[2]}
~\left\langle~l-1~m'~\vert~\vec{x}~\vert~l~m\right\rangle.\eqno(41)$$
>From parity arguments one can also write:
$$\left\langle
l~m'\left\vert~\partial_k~\right\vert~l~m\right\rangle~=~0.\eqno(42)$$
The matrix elements of $\vec{x}$ can be calculated 
(see next section) so that from
replacing the matrix elements of $\vec{\partial}$ by those
of $\vec{x}$ with the aid of eqs. (40) and (41) one can obtain
the eigenvalues of $\vec{\partial}^2$. These are:
$$\left\langle~l~m\vert~\vec{\partial}^2~\vert~l~m\right\rangle~=~
-{[2l]\over[2]}~{[2l+1]\over[2]}~-~c_l^2.\eqno(43)$$

At the begining of this section we mentioned that the component of
$\vec{p}$ paralel to $\vec{x}$ is assumed to have the form
$\vec{x}~{1\over
r}~f\left(r{\partial\over\partial r}+1\right)$. For simplicity we take here 
$f(x)=x$. In this case the realization of the 
$q$-linear momentum $\vec{p}$ reads:
$$\vec{p}~=~{-i\over r}~\left(~\vec{x}~(
r{\partial \over\partial r}~+~1)-\vec{\partial}~ \right).\eqno(44)$$
Then using equations (38) and (43) one can write:  
$$\vec{p}~^2~ {\tilde Y}_{lm}=\left[~-{1\over r}~{\partial\over\partial r}
\left(r{\partial\over\partial r}~+~1\right)~
+~{1\over r^2}\left({[2l]\over[2]}~{[2l+2]\over[2]}~+~c^2_l~-~c_l\right)
\right]{\tilde Y}_{lm}.\eqno(45)$$
One can see that in the limit $q \rightarrow 1$ one recovers the action 
of the Laplace operator on a spherical harmonic which justifies
our choice for $f$.

We mention that it is a simple but a tedious matter to calculate the
commutation relations between $\vec{r}$ and $\vec{p}$ and to verify that one
gets the right result for $q=1$. We do not display these
commutation relations
here because they are rather intricate and unnecessary in the
derivation of a covariant Schr\"odinger equation.

We also note that the operator $\vec{\Lambda}$, behaving as a vector
under the $su_q(2)$ algebra, can be written as a cross product of $\vec{r}$
and $\vec{p}$, but this does not bring any simplification because of the 
commutation relations between $\vec{r}$ and $\vec{p}$.
\vskip1cm
{\bf 5. The eigenfunctions of the {\large$q$}-angular momentum}
\vskip0.5cm
By definition, the basis vectors $\Phi_{lm}(q,x_0,\varphi)$ forming an  
invariant subspace for a $(2l+1)$-dimensional 
irreducible representation of $su_q(2)$ are 
eigenfunctions of $L_0$ and of the Casimir operator $\rm C$
of equation (4). We begin by writing them as polynomials in $x_0$
multiplied by $x_1^m$:
$$\Phi_{lm}(q,x_0,\varphi)~=~x_1^m~\sum_{k\geq0} ~a_k~x_0^k,\eqno(46)$$
where the sum runs either over $k$ even when  $l-m$  is even
or over $k$ odd when  $l-m$  is odd. 
In both cases it runs up to  $l-m$  but it starts at zero
for  $l-m$  even and at 1 for  $l-m$  odd. 

As for the $R(3)$ case, the basic equation which determines 
the matrix elements of $L_+$ and $L_-$ reads:
$$L_+~L_-~\Phi_{lm}(q,x_0,\varphi)~=~[l+m]~
[l-m+1]~\Phi_{lm}(q,x_0,\varphi), \eqno(47)$$
This equation leads to the recursion relation:
$$a_{k+2}~=~-q^{-2m}~{[l-m-k]~[l+m+k+1]\over[k+1]~[k+2]}~a_k.\eqno(48)$$
Then taking $a_0$ = 1 we obtain for  $l-m$  even:
$$\Phi_{lm}(q,x_0,\varphi)~=~x_1^m~\left\{1-{[l-m][l+m+1]\over
[2]!}~\left(q^{-m}x_0\right)^2\right.$$
$$\left.+{[l-m][l-m-2][l+m+1][l+m+3]\over
[4]!}\left(q^{-m}x_0\right)^4-...\right\},\eqno(49)$$
while for  $l-m$  odd we get:
$$\Phi_{lm}(q,x_0,\varphi)~=~x_1^m~\left\{{1\over[1]!}\left(
q^{-m}x_0\right)-{[l-m-1][l+m+2]\over
[3]!}~\left(q^{-m}x_0\right)^3\right.$$
$$\left.+{[l-m-1][l-m-3][l+m+2][l+m+4]\over
[5]!}\left(q^{-m}x_0\right)^5-...\right\}.\eqno(50)$$
In order to express these results in terms of a $q$-hypergeometric 
series it is necessary to write all the
$q$-numbers $[n]$ in the form
$$[n]~=~{q^n-q^{-n}\over
q-q^{-1}}~=~[2]~{(q^2)^{n\over2}~-~(q^2)^{-{n\over2}}\over
q^2-q^{-2}}~=~ [2]~\left[{n\over2}\right]_{q^2}.\eqno(51)$$
For  $l-m$  even we have then:
$$\Phi_{lm}(q,x_0,\varphi)~=~x_1^m~_2{\rm
F}_1\left(~q^2~;~{l+m+1\over2}~,~{-l+m\over2}~;~{1\over2}~;
~q^{-m}x_0^2~\right),\eqno(52)$$
while for  $l-m$  odd we get: 
$$\Phi_{lm}(q,x_0,\varphi)~=~x_1^m~q^{-m}~x_0~_2{\rm
F}_1\left(~q^2~;~{l+m+2\over2}~,~{-l+m+1\over2}~;~{3\over2}~;
~q^{-m}x_0^2~\right).\eqno(53)$$
The argument $q^2$ in $_2{\rm F}_1$ specifies that all the $q$-numbers
in the series expansion of $_2{\rm F}_1$ must be calculated
with $q^2$ instead of $q$.

Moreover we found that the functions $\Phi_{lm}(q,x_0,\varphi)$ 
satisfy the following 
simple relations:
$$x_1~{1\over
x_0}~{1-q^{-2N_0}\over1-q^{-2}}~\Phi_{lm}(q,x_0,\varphi)
~=~-[l-m]~[l+m+1]~\Phi_{l~m+1}(q,x_0,\varphi),\eqno(54)$$
for  $l-m$  even, and
$$x_1~{1\over
x_0}~{1-q^{-2N_0}\over1-q^{-2}}~\Phi_{lm}(q,x_0,\varphi)
~=~\Phi_{l~m+1}(q,x_0,\varphi),\eqno(55)$$
for  $l-m$  odd.

The normalized eigenfunctions of $\rm C$ and $L_0$ 
take now the form:
$$Y_{lm}(q,x_0,\varphi)=(-1)^{l-m\over2}\sqrt{[2l+1]\over4\pi}
\left({[l-m-1]!!\over[l-m]!!}~{[l+m-1]!!\over[l+m]!!}\right)^{1/2}
[2]^{m\over2}\Phi_{l~m}(q,x_0,\varphi),
\eqno(56)$$ 
for  $l-m$  even, and
$$Y_{lm}(q,x_0,\varphi)=(-1)^{l-m-1\over2}~\sqrt{[2l+1]\over4\pi}
\left({[l-m]!!\over[l-m-1]!!}{[l+m]!!\over[l+m-1]!!}\right)^{1/2}
[2]^{m\over2}\Phi_{l~m}(q,x_0,\varphi),\eqno(57)$$ 
for  $l-m$  odd. Their orthogonality relation becomes:
$$\int~Y_{l'm'}^+(q,x_0,\varphi)~Y_{lm}(q,x_0,\varphi)~d\varphi~
d[x_0]~=~ \delta_{ll'}~\delta_{mm'},\eqno(58)$$
where the integral over $\varphi$ is the same as for 
spherical harmonics, while the
integral over $d[x_0]$ defined on the interval (-1,1) is the sum of 
$$\int^1_0~x_0^n~d[x_0]~=~{1\over[n+1]}\eqno(59)$$
and of
$$\int_{-1}^0~x_0^n~d[x_0]~=~(-1)^n~{1\over[n+1]}\cdot\eqno(60)$$
The phase appearing in the right-hand side of the integral (60) is due to 
parity arguments. 
The relation (59) is in fact the result of a discrete integration
of $f(x_0)=x_0^n$, performed by dividing the integration interval
(0,1) in an infinite set of segments located
between two succesive points $x_k=q^k$ and $x_k=q^{k+1}$ where $q<1$ 
$$\int_0^{1}~f(x_0)~d[x_0]~=~
\sum_{k=0}^\infty~f(x_{2k+1})~(x_{2k}~-~x_{2k+2})~.\eqno(61)$$

Looking now for the properties of $Y_{lm}$,  
just as in the $R(3)$ case, we found that the product $x_k~Y_{lm}$ can be
expressed in terms of $Y_{l+1,~m+k}$ or $Y_{l-1,~m+k}$ as follows:
$$x_1~Y_{lm}(q,x_0,\varphi)~=~q^{l-m}~\sqrt{[l+m+1][l+m+2]
\over[2][2l+1][2l+3]} 
~Y_{l+1~m+1}(q,x_0,\varphi)$$
$$-~q^{-l-m-1}~\sqrt{[l-m][l-m-1]\over[2][2l+1][2l-1]}~ 
Y_{l-1~m+1}(q,x_0,\varphi),\eqno(62)$$
\vskip0.5cm
$$x_0~Y_{lm}~=~q^{-m}~\sqrt{[l-m+1][l+m+1]\over[2l+1][2l+3]}
~Y_{l+1~m}(q,x_0,\varphi)$$
$$~-~q^{-m}~\sqrt{[l-m][l+m]\over[2l+1][2l-1]}~
Y_{l-1~m}(q,x_0,\varphi),\eqno(63)$$
\vskip0.5cm
$$x_{-1}~Y_{lm}(q,x_0,\varphi)~=~q^{l-m}~
\sqrt{[l-m+1][l-m+2]\over[2][2l+1][2l+3]} 
~Y_{l+1~m-1}(q,x_0,\varphi)$$
$$~-~q^{l-m+1}~\sqrt{[l+m][l+m-1]\over[2][2l+1][2l-1]}~ 
Y_{l-1~m-1}(q,x_0,\varphi).\eqno(64)$$
\vskip0.5cm
In addition, we have found three relations which express the
non-commutativity of $x_k$ with $Y_{lm}$ and represent a
generalization of the equations (15) and (16):
$$x_0~Y_{lm}(q,x_0,\varphi)~=~q^{-2m}~Y_{lm}(q,x_0,\varphi)~x_0,
\eqno(65)$$ 
\vskip0.5cm
$$x_1~Y_{lm}(q,x_0,\varphi)~=~Y_{lm}(q,x_0,\varphi)~x_1$$
$$~+~{\lambda\over
\sqrt{[2]}}~q^{-m-1}~\sqrt{[l-m][l+m+1]}~Y_{l~m+1}(q,x_0,\varphi)~x_0,
\eqno(66)$$
\vskip0.5cm
$$x_{-1}~Y_{lm}(q,x_0,\varphi)~=~Y_{lm}(q,x_0,\varphi)~x_{-1}$$
$$~-~{\lambda\over
\sqrt{[2]}}~q^{-m+1}~\sqrt{[l+m][l-m+1]}~Y_{l~m-1}(q,x_0,\varphi)~x_0.
\eqno(67)$$

The last two equations have been obtained from (65) 
by acting with $L_+$ or $L_-$ which leads to a rising
or lowering of $m$ in $Y_{lm}$.

\vskip1cm
{\bf VI. A {\large$q$}-deformed Schr\" odinger equation}
\vskip0.5cm
Taking into account all the above results, we assume that the 
Hamiltonian entering the $q$-deformed Schr\"odinger equation is: 
$${\cal H}~=~{1\over2}~\vec{p}~^2~+~V(r)\eqno(68)$$
where operator $\vec{p}$ has been defined in the fourth
section.
The eigenfunctions of this
Hamiltonian are: 
$$\Psi(r,x_0,\varphi)~=~r^L~u_L(r)~Y_{lm}(q,x_0,\varphi)\eqno(69)$$
where $L$ is the
solution of the  following equation:
$$L(L+1)~=~{[2l]\over[2]}{[2l+2]\over[2]}~+~c_l^2~-~c_l\eqno(70)$$
obtained from the requirement that $u_L(r)$ remains finite in the limit
$r\to0$.

This Schr\" odinger equation has simple solutions for the Coulomb
potential
$V(r)=-r^{-1}$ and for the oscillator potential
$V(r)={1\over2}~r^2$. The eigenvalues of the two Hamiltonians are:
$$(E_{nl})_{Coulomb}~=~-{1\over2(n+L+1)^2}\eqno(71)$$
for the Coulomb potential and 
$$(E_{nl})_{oscillator}~=~(2n+L+{3\over2}).\eqno(72)$$
for the oscillator potential,
$n$ being the radial quantum number and $L$ the solution of the equation
(70), usually not an integer. We notice that the spectrum is
degenerate with respect to the magnetic quantum number $m$, i.e. the
essential degeneracy subsists.
But the eigenvalues (71) and (72) depend on two quantum numbers
so that the accidental degeneracy of the $q = 1$ case is removed. The
dependence of eigenvalues on $q$ can be obtained through solving equation
(70) for $L$. 

The solution of the wave equation which does not depend on
$\theta$ and $\varphi$ gives for the expectation value of $x_0^2$ the
value $R^2/[3]$ instead of $R^2/3$ obtained in the case of spherical
symmetry. The quantity  $R^2$ denotes the expectation value of the
operator $r^2$ in each case.
It then results that the quadrupole moment as well
as all the $2^{2n}$-poles are different from zero, although the
wave function does not depend on $\theta$ and $\varphi$.
This clearly shows that the Hamiltonian (68-70) has
lost the spherical symmetry. One can mention however that it 
gained another one, namely the symmetry under the $su_q(2)$ algebra
which may have new physical implications.

We remark that there are three sources producing differences
in the eigenvalue problem
between the case of $q$-deformed Schr\"odinger equation and the case of 
spherical symmetry. The first one is that the $q$-functions
$Y_{lm}(q,x_0,\varphi)$ 
differ from the spherical harmonics $Y_{lm}(\theta,\varphi)$ as 
shown in Section 5. The
second reason is that the coefficient of the centrifugal potential in
the radial Schr\"odinger equation is proportional 
to $L(L+1)$, with $L$ given by eq.(70),
and not to $l(l+1)$, as in the sperical case. The third source
is that in the $q$-deformed case the integral over $x_0$ is performed
according to the relations (58)-(60).

As a final comment let us recall  that for $l = 0$ one has $c_l = 1$, hence 
$L = 0$. As a consequence the $l = 0$ levels are independent of the
deformation parameter both for the harmonic oscillator and the 
Coulomb potential. An important physical aspect 
is that the centrifugal barrier disappears for $l = 0$ in 
contrast to the Hamiltonian $H_q$ of Ref. \cite{2}. Moreover the 
whole Coulomb spectrum of Ref. \cite{2} is different from ours.
It is not surprizing because this work is based on the
results of Ref. \cite{4} where 
the realization of the $su_q(2)$ generators 
and the basis
vectors of $su_q(2)$ irreps are entirely different from ours.

Physical applications with numerical examples
of the $q$-deformed Coulomb and harmonic oscillator
spectra will be considered elsewhere.\par
\vspace{2cm}
{\bf Acknowledgements}. The author is grateful to Fl. Stancu for
useful comments and a critical reading of the manuscript. He is also 
indebted to J. Beckers and Fl. Stancu for hospitality at the
University of Liege.

\end{document}